\documentclass[aps,prb,showpacs,twocolumn,superscriptaddress,unsortedaddress]{revtex4}
\usepackage{epsfig}
\usepackage{amsmath}

\newcommand{\be}{\begin{equation}}
\newcommand{\ee}{\end{equation}}
\newcommand{\bk}{{{\bf{k}}}}
\newcommand{\bq}{{{\bf{q}}}}

\newcommand{\br}{{{\bf{r}}}}

\newcommand{\bea}{\begin{eqnarray}}
\newcommand{\eea}{\end{eqnarray}}

\newcommand{\dg}{{\dagger}}
\newcommand{\pdg}{{\phantom\dagger}}
\newcommand{\nn}{\nonumber}

\begin{document}

\title{Zero Temperature  Phase Diagram of the Classical Kane-Mele-Heisenberg Model}

\author{ Mohammad H. Zare}
\author{ Farhad Fazileh} %\email{Fazileh@cc.iut.ac.ir }
\author{Farhad Shahbazi} \email{shahbazi@cc.iut.ac.ir}
\affiliation{ Department of Physics, Isfahan University of
Technology, Isfahan 84156-83111, Iran}

\begin{abstract}
The classical phase diagram of the Kane-Mele-Heisenberg model
is obtained by  three complementary methods:  
Luttinger-Tisza, variational minimization, and 
the iterative minimization method.  
Six distinct phases were obtained in the space of the couplings.  
Three phases are commensurate   with  long-range ordering,  
planar N{\'e}el states in horizontal plane (phase.I), 
planar states in the  plane  vertical to the horizontal plane (phase.VI) and
collinear states normal to the horizontal plane (phase.II). 
However the  other three, are infinitely degenerate due to the frustrating competition  between the 
couplings, and  characterized by a manifold of incommensurate wave-vectors.
These phases  are, planar helical states in horizontal plane (phase.III),
planar helical states in a vertical  plane (phase.IV) and non-coplanar states (phase.V).   
Employing the linear spin-wave analysis, it is found that the quantum fluctuations select 
a set of symmetrically equivalent states in phase.III, through the 
 quantum order-by-disorder mechanism. Based on some heuristic arguments 
is  argued  that the same  scenario may also occur in the other
two frustrated phases VI and V. 

\end{abstract}

\date{\today}

\maketitle

\section{Introduction}
The search for the Quantum spin liquid (QSL), a state preserving all 
the symmetries of a system at zero temperature,  
on the two dimensional systems  with honeycomb geometry 
has been encouraged after the quantum Monte Carlo study of Hubbard model 
on the honeycomb lattice~\cite{meng}.
This work  identified a gapped QSL state, between the Mott-insulator and semi-metal 
phases,  in a narrow interval of moderate  on-site Coulomb interactions.
 Although,  the existence of such a QSL phase was debated in  more recent 
works~\cite{debate1,debate2,debate3},  a bunch of  researches was prompted  
to study the effective spin models on the honeycomb lattice, 
arising at the strong coupling limit of the Hubbard model,
among which the  $ J_{1}-J_{2} $ model~(with positive
 nearest,$ J_1 $, and next nearest  neighbor, $ J_2 $, exchange interactions)
is the minimal spin Hamiltonian~\cite{kawamura,aron2010,sd2,
mosadeq,pvb2,pvb3,pvb4,pvb5,pvb6,pvb7,qsl1,qsl2,qsl3,qsl4,qsl5,qsl6,qsl7,eps,ring1,ring2}.  
Being bipartite, the N{\'e}el ordered ground state 
 is stable for small values of  $ J_2 $ on the honeycomb lattice.
However, it has been shown that for  $ S=1/2 $, 
 the staggered magnetization is  reduced by about fifty percent with respect to
the classical value, because of  the enhanced  
quantum fluctuations  due to the small coordination number of the lattice\cite{n1,n2,n4}. 
The next nearest neighbor interaction, $ J_2 $, tends to 
destabilize the N{\'e}el ordering
and destroys  it at a critical value $ J_2/J_1\sim 0.2 $. 
Various proposals    have been put forward for 
the nature of the disordered phase, such as 
  staggered dimerized~\cite{aron2010,sd2}  
(a valence bond crystal state breaking the rotational symmetry of the lattice ),
 plaquette valence bond crystal~\cite{mosadeq,pvb2,pvb3,pvb4,pvb5,
pvb6,pvb7} 
(a resonating valence bond state breaking the translational symmetry of the lattice), 
and quantum spin liquid ~\cite{qsl1,qsl2,qsl3,qsl4,qsl5,qsl6,qsl7}. 

The theoretical and experimental achievements  in the field 
of topological insulators (TI)~\cite{TI1,TI2}, attracts many
attentions for exploring the effect of electron-electron interaction
in the  systems with strong spin-orbit
coupling, where the topological features arises~\cite{TI-int}. 
One of the earliest model
in this context, introduced  by Kane and Mele (KM), which is  a tight-binding
model of non-interacting electrons on a honeycomb lattice subjected  to 
a spin-orbit term~\cite{Kane2005}.  Electron-electron interaction can be 
introduced to the KM model by adding an on-site repulsive Hubbard term
to this model, resulting in a Hamiltonian,  so-called  the Kane-Mele-Hubbard model. 
Analytic as well as the numeric study of this model, suggest a QSL phase
at the moderate Hubbard interaction, between the topological band insulator (TBI)
and Mott insulator phases~\cite{Rachel2010,kmh1,kmh2,
kmh3,kmh4,kmh5,kmh6,kmh7,kmh8,kmh9}.
 The strong coupling limit ($ U\gg t $)
of the model can be described, effectively, by a XXZ spin Hamiltonian
namely, the Kane-Mele-Heisenberg  (KMH) model~\cite{Rachel2010,vaezi}. 
Recently, Vaezi et al~\cite{vaezi}, using the Schwinger-boson and Schwinger-fermion 
methods,  proposed 
a chiral spin liquid state for this model,  in a narrow region of  the phase diagram,     
the rest of which is divided into the N{\'e}el and incommensurate N{\'e}el ordered phases.   

A promising  guideline  of  finding the possibility of the spin liquid state for a 
spin Hamiltonian, 
is exploring of its  classical phase diagram (the large-$ S $ limit), at zero temperature.
The regions in the 
phase diagram where the classical ground state is highly degenerate, due to
the competition between  the couplings, are likely to be QSL in the quantum 
limit.  The classical phase diagram of the Anti-ferromagnetic  Heisenberg model
on the honeycomb lattice, up to third neighbor interaction, has been 
extensively studied earlier~\cite{Rastelli, Fouet}, and various ordered 
phases as well as  a region of magnetically disordered phase has been obtained.    

The absence of such an analysis for the KMH model was our motivation for this work.
We use  Luttinger-Tisza, Variational and Iterative minimization methods,
each being complementary of the other, to obtain the  classical phase diagram of 
the KMH model. The rest of paper is organized as the following.   
In Sec.II, the KMH model is introduced, Sec.III is devoted to 
the extraction  of  the classical phase diagram by the three mentioned methods.
Using linear spin wave theory, we  calculate the  quantum correction 
in Sec.IV, and investigate the stability of the classical phase as well as the 
possibility of quantum order-by-disorder in the classically degenerate regions.  
finally, we summarize   the results in Sec.V.

\section{Kane-Mele-Heisenberg Model}
Kane and Mele~\cite{Kane2005} proposed a model for describing  the quantum spin Hall (QSH) effect in graphene, 
by adding a mass term to the nearest neighbor tight-binding Hamiltonian. The model is defined by the following Hamiltonian
\be
H_{\textbf {KM}}=-t\sum_{\langle ij\rangle,\sigma}c_{i\alpha}^\dagger c_{j\alpha}
+i\lambda\sum_{\langle\!\langle i,j\rangle\!\rangle,\alpha\beta}\nu_{ij}\sigma_{\alpha\beta}^z c_{i\alpha}^\dagger c_{j\beta},
\label{kane2.eqn}
\ee
in which the first term represents the hopping 
between the  nearest neighbors of a honeycomb lattice,   
and the second  term , with $ \nu_{ij}=\pm 1 $ being an anti-symmetric tensor, 
denotes the  
next  nearest neighbor spin-orbit hopping integral which  gives rise 
 to the topological features, such as the  edge state modes.

Adding   electron-electron interaction  to the 
Kane-Mele Hamiltonian, Eq.~(\ref{kane2.eqn}),  by 
 an on-site repulsive Hubbard term
\be
H_{U}=\sum_{i}U n_{i\uparrow}n_{i\downarrow},
\label{HU.eqn}
\ee
gives the Kane-Mele-Hubbard model.   
In the limit where the on-site coulomb repulsion $U$ is much larger than the hopping integrals  
$t$ and $ \lambda $,  the charge fluctuations
are suppressed, and one can project the half-filled Hamiltonian to the
 lowest Hubbard sub-band for which the condition of the single-occupied site is fulfilled. 
 This procedure can be done perturbatively in terms of the ratios $t/U$ and $\lambda/U$.  To the fourth order,  
 we find  a spin Hamiltonian called the 
 Kane-Mele-Heisenberg ($ {\textbf {KMH}} $) model  
 as following\cite{Rachel2010,vaezi}         

\begin{equation}
\begin{split}
  H_{\textbf {KMH}}\!\!&=\!\!J_1\sum_{\langle i,j\rangle}{\bf S}_{i}\cdot {\bf S}_{j}
   +J_2\sum_{\langle\!\langle i,j\rangle\!\rangle}{\bf S}_{i}\cdot {\bf S}_{j}\\
   &+g_2\sum_{\langle\!\langle i,j\rangle\!\rangle}(-S_{i}^x~S_{j}^x-S_{i}^y~S_{j}^y+S_{i}^z~S_{j}^z),
\end{split}
\label{kmh}
\end{equation}
in which $J_{1}=4t^2/U-16{t^4}/U^3$, $J_{2}=4{t^4}/U^3$
  and $g_{2}=4{\lambda^2}/U$ 
are the first and second neighbor exchange couplings and 
$ {\bf S}_i  $ represents an $ S=1/2 $ spin residing in site $ i $. 
We are interested in classical phase diagram of this model, hence
in the  large $ S $ limit , we consider the spins as unit  vectors.
In the next section we use three methods 
to find the classical phase diagram at the zero temperature.   

\section{Classical phase diagram of KMH model}  

\subsection{Luttinger-Tisza(LT) Method}

\begin{figure}[t]
\includegraphics[trim=0cm 0cm 0cm 0cm, clip, scale=0.3]{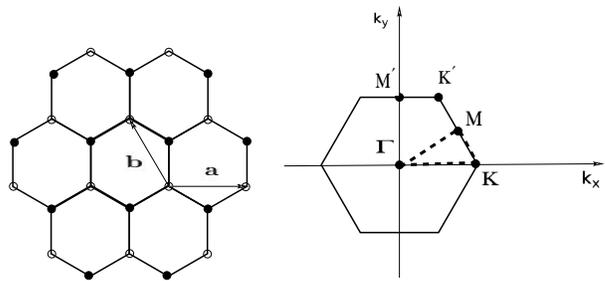}
\caption{ Left: geometry of the real space honeycomb lattice. Primitive vectors are denoted by $ {\bf a} $ and 
$ {\bf b}  $, and the open and filled circles  denote the two triangular sub-lattice points. Right:The  first Brillouin zone of the honeycomb lattice.    }
\label{honeycomb}
\end{figure}

The LT method  \cite{LT,Kaplan1960,Kaplan2006} 
 is a way of  finding the ground state  of a classical quadratic Hamiltonian. 
By Fourier transformation of the spins and the couplings,  one can  find  a matrix representation for a quadratic  spin Hamiltonian in the Fourier space. The lowest eigenvalues and eigenvectors
 of this matrix  give the ground state energy and the corresponding spin structure factor, respectively.     

Honeycomb lattice is composed of the two triangular sub-lattices~(Fig.~\ref{honeycomb}),  therefore 
the Fourier transformations of the spins for each sub-lattice become:

\begin{equation}
\begin{split}
\textbf{S}^1({\bf r}_ i) = \frac{1}{\sqrt{N/2}}\sum_{\bf q} \textbf{S}^1_{\bf q} e^{-i\textbf{q} \cdot \textbf{r}_i},\\    
\textbf{S}^2({\bf r}_ i) = \frac{1}{\sqrt{N/2}}\sum_{\bf q} \textbf{S}^2_{\bf q} e^{-i\textbf{q} \cdot \textbf{r}_i},
\label{kmhamilton:eq}
\end{split}
\end{equation}
where $ \textbf{r}_{i} $  denote  the translational vectors of the 
triangular Bravais lattice  and $ N/2 $ is the number of
 primitive cells, and the superscripts denote the sub-lattices. 
 Substituting the above transformations into  
the classical $\textbf{KMH}$ Hamiltonian, Eq.(\ref{kmh}), one gets  
the following form for the Hamiltonian in terms of the Fourier components:
\begin{equation}
H^{\rm cl}_{\textbf{KMH}} = \sum_{\bq} {\bf S}^{t}_{-\bq}\cdot {\bf M}_{\bq}\cdot {\bf S}_{\bq},
\label{kmh.q}
\end{equation}
in which 
${\bf S}^{t}_{-\bq} = 
\begin{pmatrix}S_{-{\bq},x}^1 \ S_{-\bq,y}^1 \ S_{-\bq,z}^1 \ S_{-\bq,x}^2 \ S_{-\bq,y}^2 \ S_{-\bq,z}^2 \end{pmatrix}$,
and the matrix ${\bf M}_{\bq}$ is  defined as
\be
{\bf M}_q = \begin{pmatrix} A_\bq & 0 & 0 & C_\bq & 0 & 0  \\
0 & A_\bq & 0 & 0 & C_\bq & 0  \\ 
0 & 0 & B_\bq & 0 & 0 & C_\bq  \\
C^*_\bq & 0 & 0 & A_\bq & 0 & 0 \\
0 & C^*_\bq & 0 & 0 & A_\bq & 0 \\
0 & 0 & C^*_\bq & 0 & 0 & B_\bq 
\label{Mk1}
\end{pmatrix}
\ee
whose elements are given by
\bea
A_\bq &=& 2(J_{2}\!-\!g_{2}) [ \cos q_{a} \!+\! \cos q_{b} \!+\! \cos(q_{a} \!+\! q_{b}) ]\nn \\
B_\bq \!&\!=\!&\! 2(J_{2}\!+\!g_{2})[\cos q_{a}+\cos q_{b}+\cos(q_{a} + q_{b})] \nn \\
C_\bq &=& J_1 [ 1 \!+\! \exp({iq_b}) +\exp({i(q_a+q_b)}) ].
\label{ElementsMK}
\eea

Here $ q_a={\bf  q}\cdot{\bf a}  $ and $q_b={\bf  q}\cdot{\bf b}  $,
 where ${\bf a}=\hat{\bf x}$ and ${\bf b}=-1/2~\hat{\bf x}+{\sqrt{3}}/{2}~\hat{\bf y}$ are the primitive translational  
vectors of the honeycomb lattice displayed in Fig.~\ref{honeycomb}. 
Writing Eq.(\ref{kmh.q}) in terms of the  eigen-modes of ${\bf M}_{\bq}$, 
leads us to the following simple quadratic form
\be
H^{\rm cl}_{\textbf{KMH}}=\sum_{\mu}\sum_{\bq} 
 \lambda_{\bq}^{\mu} |{\tilde{\bf S}}_{\bq}^{\mu}|^{2}
\label{ECL}
\ee
where $ \lambda_{\bq}^{\mu} $  with $ \mu=1,2,3,4,5,6 $ are the eigenvalues of ${\bf M}_{\bq}$, and 
\be
{\tilde{\bf S}}_{\bq}^{\mu}= w^{\mu}_{\bq}{\bf S}_{\bq},
\ee
denotes the spin structure factors, with $ w^{\mu}_{\bq}  $ being the eigenvectors of 
 ${\bf M}_{\bq}$ with eigenvalue $ \lambda_{\bq}^{\mu}  $. 
\be
{\bf M}_{\bq} w^{\mu}_{\bq} =\lambda_{\bq}^{\mu}w^{\mu}_{\bq}. 
\ee

The  size of spins being unity at each site,  
requires the  following constraint be satisfied by the Fourier components 
\bea
\sum_{i} | {\bf S}_{i}|^{2}=\sum_{\bq} |\tilde{\bf S}_{\bq}|^{2}=N.
\label{const}
\eea
If a global minimum $ \lambda_{0} $ is found among  the eigenvalues, 
using Eq.(\ref{const}) the classical energy (Eq.(\ref{ECL})) can be rewritten  as
\be
H^{\rm cl}_{\textbf{KMH}}=N\lambda_{0}+\sum_{\mu>0}\sum_{\bq} 
 (\lambda_{\bq}^{\mu}-\lambda_{0}) |{\tilde{\bf S}}_{\bq}^{\mu}|^{2}.
\ee
Since $ \lambda_{\bq}^{\mu}-\lambda_{0}>0 $,  
the ground state energy per primitive cell is equal to $ \lambda_{0} $,
 provided that all the structure factors other than the
 one corresponding to the minimum eigenvalue vanishes. 
 These conditions enables us to find the  spin configuration in the ground state, as well.   

For the $ \textbf{KMH} $ model, the eigenvalues of   $M_{\bq}$  are given by
\begin{eqnarray}
 \lambda_{1,2}= A_{\bf q}-\vert{C_{\bf q}\vert}, \nn \\
 \lambda_{3,4}= A_{\bf q}+\vert{C_{\bf q}\vert}, \nn \\
 \lambda_{5}= B_{\bf q}-\vert{C_{\bf q}\vert}, \nn \\
  \lambda_{6}= B_{\bf q}+\vert{C_{\bf q}\vert},
\end{eqnarray}
where $ A_{\bf q}, B_{\bf q} $ and $ C_{\bf q} $ are given by Eq.(\ref{ElementsMK}).
The two-fold degeneracy of the first two eigenvalues reflects the $ O_{2} $ symmetry of the ${\bf KMH}$ model in $x-y$ plane.   

 \begin{figure}[t]
\includegraphics[trim=2.5cm 0cm 0cm 0cm, clip,  scale=0.45]{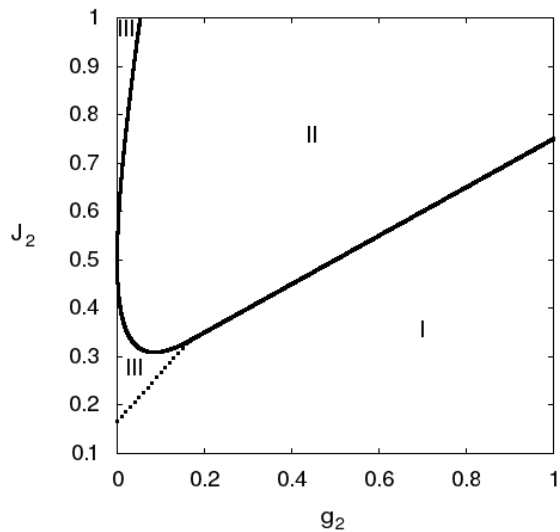}
\includegraphics[scale=0.3]{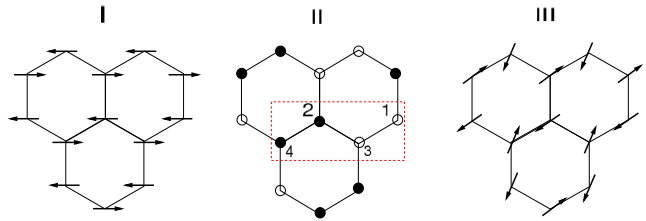}
\caption{Top: The Luttinger-Tisza phase diagram of the
 $ {\bf  {KMH}}$ Hamiltonian  in $J_{2}-g_{2}$ space. 
 Solid and dotted curves denote the first-order and secon-order 
 phase boundaries, respectively.   
 Bottom:  typical spin configuration in  I)  N{\'e}el-$ xy $,
 II) collinear-$ z $  state  with the wave-vector $ {\bf M}=(\frac{\pi}{a}, \frac{\pi}{\sqrt{3}{a}}) $ 
 (open and filled  circles  denote the up and down spin directions, respectively) and 
 III)  helical-$ xy $ phase.     }
 \label{LT-phase}
\end{figure}

Setting $ J_1=1 $, we proceed to find the stable phases of the $ {\bf KMH} $ model by 
calculating  the minimum eigenvalues of the coupling matrix ,
  for  $ 0<J_{2}<1 $ and $0< g_{2}<1$.  
Fig.~\ref{LT-phase} represents the phase diagram obtained by LT method, 
departing  the coupling  space into the   three distinguished  regions:

I )In-plane commensurate  N{\' e}el  states~(N{\'e}el-$ xy $))-
the minimum eigenvalues indicating this phase are $ \lambda_{1,2}, $ 
with the wave-vector  $q=(0,0)$ and phase difference $ \pi $ between the two spins within each unit-cell. 
The  $ O_{3} $ symmetry of  Heisenberg Hamiltonian is reduced to $ O_2 $ due to the  $g_{2}$ term, 
laying  the spins in the  $xy$-plane. The energy per spin in this phase is given by: 
 \bea
  %\bq_{x}=0, \bq_{y}=, \nn \\
  \epsilon_{I}=3(J_2-g_2)-\frac{3}{2}.
   \eea
 
 II) Commensurate vertical collinear  states   ( collinear-$ z $): 
 this phase is three-fold degenerate and characterized by   alignment of spins  along the ${\bf \hat{z}} $ axis, normal to the honeycomb plane. 
 In this phase the eigenvalue $ \lambda_{5} $ is minimized  by 
 the following three wave-vectors ($ M $-points in  $1BZ$)   
 \bea
  && q_x=\frac{\pi}{a}, q_y=\pm\frac{\pi}{\sqrt{3}{a}},\nn\\
  && q_x=0, q_y=\frac{2\pi}{\sqrt{3}{a}},
 \eea
 and the phase difference $ \pi $ within each unit-cell. The energy per spin is given by
 \be 
 \epsilon_{II}=-(J_2+g_2)-\frac{1}{2}. 
 \ee

The  boundary between this phase and the  phase.I,  is given by    
\be
J_2=\frac{g_2}{2}+\frac{1}{4},~g_{2}> 1/6 
\ee
 which is a first-order transition line. \\

 III) Incommensurate in-plane helical  states (helical-$xy$):
in this phase,  the eigenvalues $ \lambda_{1,2} $ become minimum at the incommensurate wave-vector 
\be
q_x=\frac{2}{a}\cos^{-1}\left({\frac{1-2(J_2-g_2)}{4(J_2-g_2)}}\right), q_y=0,
\ee
and give the energy per spin as 
\be
\epsilon_{III}=\frac{-1}{8(J_2-g_2)}[1+12{(J_2-g_2)}^2].
 \ee
The line
\be
J_{2}=g_2+\frac{1}{6},~~g_{2}<1/6,
\ee
is the continuous  transition boundary, separating  this phase from the phase.I, and  the  curves
\bea
J_2=3{g_2}+\frac{1}{2}-\sqrt{4{g_2}^2+2g_2}, ~~g_{2}< 1/6, J_{2}<1/2\nn \\
J_2=3{g_2}+\frac{1}{2}+\sqrt{4{g_2}^2+2g_2}, ~~ g_{2}< 1/6, J_{2}>1/2.
\eea
determine the borders  of  phase.III  and  the phase.II, which are first-order boundaries.     

The problem with the LT method is that the ground states derived by this are only 
consistent with the weak constraint Eq.(\ref{const}), hence there is no 
 guarantee that the spin configurations with minimum energy satisfy the 
strong constraint of the unit  spin size  at each site. 
Indeed,  it gives an upper bound for the classical ground state energy, 
 Moreover the non-coplanar spin configurations can not be found by this method~\cite{Henley1,Henley2}.
 To search for such states, 
we parametrize each   spin  in terms of its wave-vector, 
 polar and  azimuthal angles and use variational  minimization  method to find 
 the ground state.         

\subsection{Variational minimization  Method}

 In this section we parametrize the spins in such a way 
 that the constraint of the unity of spin size is fulfilled at each 
 lattice site. We start to search for planar states. 
 Consider a  planar-$ xy $ pattern 
 with wave-vector $ {\bf q} $,  the spins in this configuration 
 can be  parameterized as 
 
 \bea
 S^{1}({\bf r}_{i})&=&S[\cos({\bf q}.{\bf r}_{i})
 \hat{\bf x}+\sin({\bf q}.{\bf r}_{i})\hat{\bf y}]\nn\\
 S^{2}({\bf r}_{i})&=&-S[\cos({\bf q}.{\bf r}_{i}+\varphi)
 \hat{\bf x}+\sin({\bf q}.{\bf r}_{i}+\varphi)\hat{\bf y}],
 \label{hel}
 \eea
 in which  $ \varphi+\pi $ denotes the phase difference between the spins in a primitive cell. 
Substituting  Eq.(\ref{hel}) in the
 $ {\bf KMH} $ Hamiltonian (Eq.(\ref{kmh})),
 we get the following expression for the classical energy per spin
\bea
\epsilon_{\rm cl} &=&-\frac{J_1S^2}{2} [ \cos \phi + \cos(\phi -q_b) + \cos(\phi - q_a - q_b) ]  \nn \\
&+& (J_2-g_2) S^2[ \cos q_a+\cos q_b + \cos(q_a+q_b)].
\label{eq:EclLT}
\eea
Minimization of  this energy with respect to $q_x$ and $q_y$  and $ \varphi $, gives
$ q_x=q_y=0 $ and $\varphi=0  $ in the phase.I resulting in the in-plane  N{\'e}el states.  

\begin{figure}[t]
\includegraphics[trim=8cm 2cm 0cm 5cm ,clip, scale=0.45]{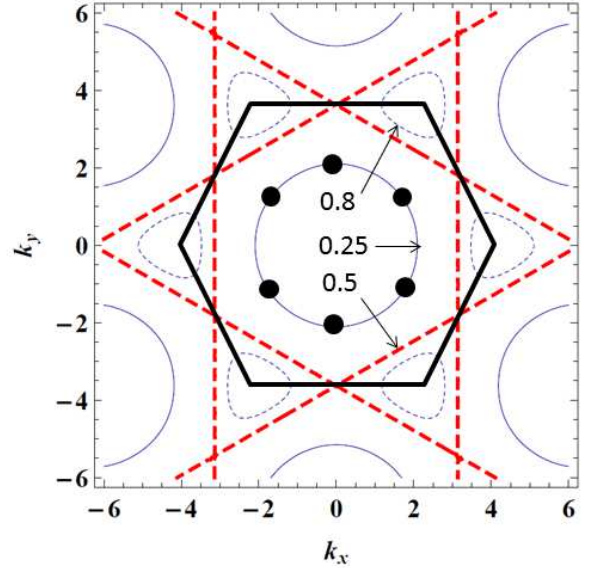}
\caption{The countours of  classical  ground state. The numbers denote the
values of $ J_{2}-g_{2} $. }
\label{fig:frustrate}
\end{figure}

For the phase.III we get
 the following relations 
\bea
&&\cos q^*_a + \cos q^*_b + \cos (q^*_a + q^*_b) = 
\frac{1}{2} \left[\left(\frac{1}{2(J_2-g_2)}\right)^2-3\right],\nn\\
&&\sin\phi^* = 2 {(J_2-g_2)} \left(\sin q^*_b + \sin(q^*_a + q^*_b)\right),\nn\\
&&\cos\phi^* = 2 (J_2-g_2) (1+\cos q^*_b + \cos (q^*_a+q^*_b)),
\label{eq:spiralJG}
\eea
meaning  the  infinitely degenerate set of in-plane helical states, 
which form a manifold for the ground state.  
Similar results have already been obtained for $ J_{1}-J_{2} $ model ($ g_{2}=0 $) in Ref. \cite{aron2010}.
Fig.~\ref{fig:frustrate} displays the ground state manifold in 
the first Brillouin zone. This  is in the form a closed countour around  $q=(0,0)$ for $1/6<(J_2-g_2)<1/2$, 
while it encircles the $K$-points  for $(J_2-g_2)>1/2$.

To search for  the non-coplanar ground states,  
one  needs to include all  the three component of the spins. 
In the simplest case,  the minimum magnetic unit cell of a spin pattern 
consists of two spins,  therefore the spins can be parametrized as the following 

\begin{figure}[t]
\includegraphics[trim=2.5cm 1cm 0cm 0cm, clip, scale=0.5]{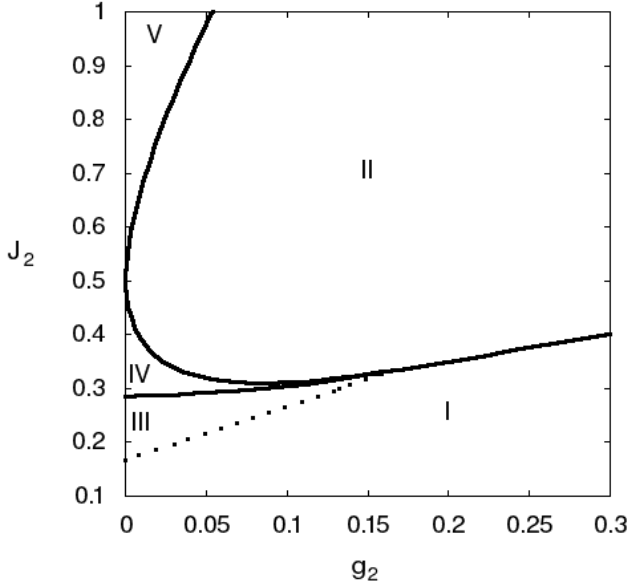}
\caption{The phase diagram obtained by variational minimization method. 
This method divides  the phase.III, previously obtained by LT method,
 into  three regions. 
The helical-$ xy $ states remain in the region assigned with the phase.III, 
 the  incommensurate planar states in a plane vertical  to $ xy $-plane (phase.IV), and 
the incommensurate non-coplanar states (phase.V).}
\label{fig:PMphase}
\end{figure}

\bea
 {S^{\alpha}_{i,x}} &=& S\sin(\theta^{\alpha}_{i})\cos(\phi^{\alpha}_{i}), \nn \\
 {S^{\alpha}_{i,y}} &= & S\sin(\theta^{\alpha}_{i})\sin(\phi^{\alpha}_{i}), \nn \\
 {S^{\alpha}_{i,z}} &=& S\cos(\theta^{\alpha}_{i}), \nn \\
\label{n-coplan}
\eea
in which $\alpha=1, 2$,  labels the two sub-lattices and 
\bea
\phi^{1}_{i}&=&{\bf q}\cdot {\bf r}_i,\nn\\
\phi^{2}_{i}&=&{\bf q}\cdot {\bf r}_i+\varphi,\nn\\
\theta^{1}_{i}&=&{\bf q}'\cdot {\bf r}_i+\gamma,\nn\\
\theta^{2}_{i}&=&{\bf q}'\cdot {\bf r}_i+\eta .
\eea
Substituting  the spins by Eq.(\ref{n-coplan})
 in the $ {\bf  KMH} $ Hamiltonian, minimization   of 
the classical  energy with respect to the   $ 7 $ variables 
 $q_x, q_y, {\acute{q}}_x, {\acute{q}}_y, \gamma, \eta, \phi $ 
would give us the ground state energy as well as the corresponding spin configuration. 
The simulated  annealing scheme  was used, from the 
Mathematica optimization package \cite{Wolfram},
 for the numerical minimization of  the energy function. 
The  N{\'e}el-$ xy $ (phase.I) and the collinear-$ z $ (phase.II) were easily
reproduced by finding $ ({\bf q}'=0, \gamma=\eta=\pi/2 )$ 
for the former and 
$( {\bf q}'=M,M',M-M', \gamma=0, \eta=\pi) $
for the later. 
The case of interest is the helical-$ xy $  (phase.III), for which  the 
energy per spin can be simplified as the following       	

\begin{equation}
\begin{split}
\frac{\epsilon_{cl}}{S^2} &=-\frac{J_{1}}{4}\sum_{\delta_1}[\cos(\acute{\bq}.\delta_1+\gamma-\eta)(1+\cos(\bq.\delta_1-\phi))]\\
&+\frac{(J_2-g_2)}{2}\sum_{\delta_2}[\cos(\bq.\delta_2)\cos(\acute{\bq}.\delta_2)]\\
&+\frac{(J_2+g_2)}{2}\sum_{\delta_2}\cos(\acute{\bq}.\delta_2).
\end{split}
\label{EIC.eqn}
\end{equation} 

where 
\be
\delta_1={\bf 0}, \pm\textbf{b}, \pm(\textbf{a+b})
\label{delta1}
\ee
denote  the unit cell position vectors  of the  nearest neighbors  
of  a  given lattice point  with $ - $ and $ + $  correspond to $ 1 $ and $ 2 $ sub-lattices, respectively, 
and 
\be
\delta_2=\pm\textbf{a}, \pm\textbf{b}, \pm(\textbf{a+b}),
\label{delta2}
\ee
are the position vectors  of the 
 second neighbor primitive cells. As shown in Fig.~\ref{fig:PMphase},
 Minimization of this energy tends to the partitioning of  the phase.III of  
the LT phase diagrams into three regions. \\

i) Incommensurate helical-$ xy $ as
 previously found by LT method and so we call it again phase.III, \\

ii)  Incommensurate planar phase in a plane perpendicular  to honeycomb plane (phase.IV).
Like the helical-$ xy $, this phase is also infinitely degenerate and   
the minimum energy is achieved by  $ ({\bf q}=0, \varphi=\pi) $ and a
 set of wave-vectors $ {\bf q}' $ and phase differences 
 satisfying  the following relations

\bea
&&\cos q'^*_a + \cos q'^*_b + \cos (q'^*_a + q'^*_b) = 
\frac{1}{2} \left[\left(\frac{1}{2J_2}\right)^2-3\right],\nn\\
&&\sin(\eta^*-\gamma^*) = 2 J_2 \left[\sin q'^*_b + \sin(q'^*_a + q'^*_b)\right],\nn\\
&&\cos(\eta^*-\gamma^*) = 2 J_2 [1+\cos q'^*_b + \cos (q'^*_a+q'^*_b)]
\label{eq:spiralJG2}
\eea
These relations are the same as Eq.(\ref{eq:spiralJG}), except the absence of   the coupling 
$ g_2 $. The reason that $ g_2 $ is omitted from the classical energy
in this phase can be explained as the following. 
Because of the in-plane $ O_2 $ 
symmetry, one can consider the spins in this state in the  $ xz $ plane.  Thus  substituting 
the $ x $ and $ z $ components of the spin from Eq.(\ref{n-coplan}) to the $ {\bf {KMH}} $
Hamiltonian, one finds that the contribution of the $ g_2 $ term to the energy is as 

\be
g_2\sum_{\ll i,j\gg}2\cos({\bf q}'\cdot(\br_i+\br_j)+2\gamma).
\ee
Since $ {\bf q}'$  is not equal to any of the $ M $-points in the $ 1BZ $, then 
$ 2{\bf q}' \cdot(\br_i+\br_j)\neq  2n\pi$ and so the above summation vanishes. 
Therefore, all over 
the phase.IV the energy is independent of $ g_2 $. 
However, the effect of this term is the breaking of the 
$ O_3 $ symmetry of the $ J_{1}-J_{2} $ Heisenberg Hamiltonian, 
in such a way that the spins confine  in a vertical plane. 
For  small values of $ J_2 $, the $ g_2 $ term cooperates with the $ J_1 $ term to 
ferro-magnetically align the second neighbors in the $ xy $-plane, resulting the 
in-plane N{\'e}el ordering. 
On the other hand, the term $ J_2 $ tends to 
align the second neighbor spins anti-parallel,
hence increasing $ J_2 $ to above a critical value,   the in-plane states 
become destabilized and system gains energy by lifting the spins
out of the honeycomb plane  and arrange them in a plane perpendicular to it.
It is also found that  even for $ g_2=0 $, the co-planar states found to be less energetic than 
non-coplanar ones, though with a very small  energy difference. \\ 

 iii) the incommensurate non-coplanar phase, called phase.V. This phase
is highly degenerate in terms of both
the $ xy $ and $ z $ wave-vectors, 
however  they are not equal in general ($ {\bf q}\neq{\bf q}' $ ). 
 These states are approximately planar normal to 
$ xy $-plane with a small out of plane spin component. The energy landscape in this phase 
 is very flat with a weak dependence  on $ g_2 $.         

Here, we limit ourselves  to  the states with two-spins in a unit cell.
Extending this method to states with  larger magnetic  cells is straight forward, 
however increasing  the number of 
variational parameters  makes the finding of the global
minimum extremely difficult. Hence to search for such a new state, in the next part we employ 
a method  called iterative minimization.

\subsection{Iterative Minimization Method}

For extracting the classical phase diagram for $ {\bf KMH} $ model we also apply the iterative minimization method \cite{Walker,Henley1,Henley2}. In this method we start from an initial state with spins arranged in a random configuration. Then in each step of the iterative process a random spin is selected for adjusting its orientation in order to minimize its energy. The minimization is achieved by aligning the selected spin with the local field produced by its neighbors while keeping its length unity. For the 
$ {\bf  KMH} $ model the local field in the position of spin $\textbf{S}_i$ is: 
\begin{equation}
\begin{split}
  \textbf{M}_i &= J_1\sum_{j:\langle i,j\rangle}\textbf{S}_j
   +J_2\sum_{j:\langle\!\langle i,j\rangle\!\rangle}\textbf{S}_j\\
   &+g_2\sum_{j:\langle\!\langle i,j\rangle\!\rangle}(-S_{j}^x\hat{\textbf{x}}-S_{j}^y\hat{\textbf{y}}+S_{j}^z\hat{\textbf{z}}),
\end{split}
\label{kmhamilton:eqn}
\end{equation}
where sums are run over $j$s that are nearest or next nearest neighbors of $i$ site. To minimize energy in each step we adjust spin $\textbf{S}_i$ as $\textbf{S}_i = -\textbf{M}_i/|\textbf{M}_i|$. The iterations are continued until the method converges to some (local) energy minimum.

We start with a honeycomb lattice consisting of two triangular sub-lattices of parallelogram shapes with periodic and open boundary conditions and with sizes of $8\times8$, $16\times16$, $32\times32$ and $64\times64$ sites each in total consisting of $N_s$ sites; i.e. $N_s$ is 128, 512, 2048 and 8192 respectively. Then we run a large loop, and on each iteration of the loop we pick $N_s$ spins for updating. 
 
 \begin{figure}[t]
\includegraphics[width=3.20in]{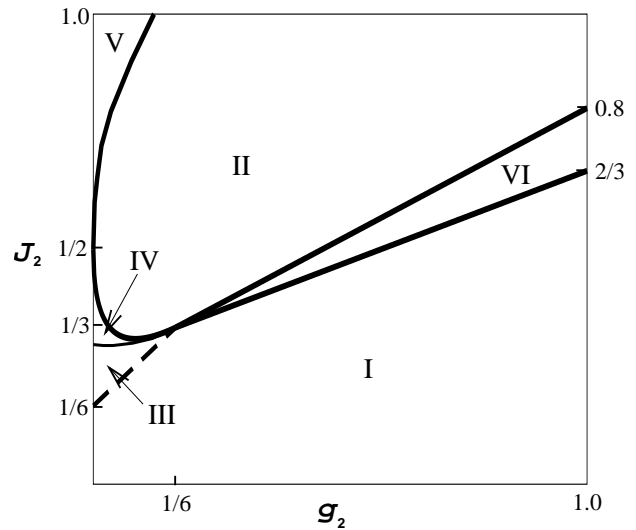}
\caption{Classical phase diagram for $ {\bf KMH} $ model obtained from Iterative minimization method.
The  newly found phase, called phase.VI,  between  phase.I and 
phase.II, is a commensurate-planar state in a  vertical plane.}
\label{fig:IMphase}
\end{figure}

\begin{figure}[t]
\includegraphics[width=3.20in]{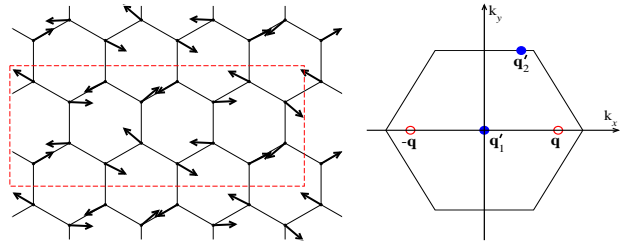}
\caption{Left: spin configuration in the commensurate  coplanar-$xz$ region  (phase.VI ).
The dashed rectangle shows  the  magnetic unit cell, consisting of $ 16  $ spins.   
 In this plot the $z$ components have been projected along the 
 $ y $-axis in order to  show the 
 spin configuration in the $xy$-plane.
  Right: the wave-vectors corresponding to the 
Fourier transformation of the spin configuration in phase.VI, in which 
open and filled circles denote the in plane  ($ {\bq}'$) and normal to plane 
($ {\bq} $) wave-vectors, respectively.  }
\label{fig:xzspin}
\end{figure}
 
Once we have obtained a local energy minimum spin configuration, we try to get a sense of the spin configuration by plotting the spin arrangements in real space and looking at their $x$, $y$ or $z$ components. For some cases that the ground state is a commensurate state, the local energy minimum state consists of finite domains similar to the true ground state, separated by domain walls. By restarting the program with a random spin configuration that is close to the proposed true ground state configuration, the program rapidly converges to the true ground state and by comparing its energy we can verify our guess for the ground state. For the regions of the phase diagram that the true ground state is a state commensurate with lattice periodicity, we start from a properly selected random configuration and normally we obtain the true ground state for small size systems with very fast convergence.  

We can also calculate the Fourier transforms of the spins to get an idea of the spin configuration. This method is suitable for both commensurate and incommensurate states. After calculating the Fourier transforms of spin components we plot the distribution of the Fourier magnitudes of spin components in the first Brillouin zone. Normally, we observe one or a few peaks in the Brillouin zone, which gives the wave vector components of the state.    

The classical phase diagram obtained from this method is shown in Fig.~\ref{fig:IMphase}.  For small  values of $J_2$ the system shows N\'{e}el ordering in  $xy$-plane 
(Phase.I)  with zero wave-vector.
 For $ g_2> 1/6 $,  increasing the  value of $J_2$ the system goes to a co-linear $z$ state
which has three-fold degeneracy  with wave-vector one of $ M $-points (phase.II).
 However, At the border phase.I and phase.II
  we found a new narrow region with  commensurate co-planar spin state
  in a plane   perpendicular to the $xy$- plane (here we call it
coplanar-$ xz $ or phase.VI). One realization of  the spin configurations in this region is shown in Fig.~\ref{fig:xzspin}.
Unlike, the helical states the in-plane and normal to plane spin components have different wave-vectors in reciprocal space. 
The $z$ spin component has two Fourier components with the wave-vectors   ${\bf q}=\pm(\pi,0)$,
while the in-plane spin component is a linear combination of two Fourier components with the wave-vectors 
${\bf q}'_{1}=(0,0)$ and ${\bf q}'_{2}=(\frac{\pi}{2},\frac{2\pi}{\sqrt{3}})$ (Fig.~\ref{fig:xzspin}).  
On the other hand the phase difference between the 
$ x $-components in each unit cell is $ \pi $, while the $ z $-components 
are in the same phase, thus the spin components in this phase can be written as 
\bea
&&S^{1}_x({\br})=-S^{2}_x({\br})=B_0+B_1\cos({\bq}'\cdot{\br} + {\phi}')\nn\\
&&S^{1}_z({\br})=S^{2}_z({\br})=A\cos({\bq}\cdot{\br + \phi}).
\eea 

For  $g_2 < 1/6$,  we observe several coplanar and non-coplanar incommensurate states.  Close to $g_2 = 0$ axis we found two degenerate incommensurate state phase regions. Our results reproduce the results of Mulder et al \cite{aron2010} for $g_2 = 0$. In the absence of $g_2$ for $\frac{1}{6} < J_2/J_1 < \frac{1}{2}$ there is a circular manifold of classically degenerate spiral state wave vectors centered at origin. We found that by turning on $g_2$, the spiral state remains stable but it becomes non-coplanar until $g_2$ exceeds $J_2 - \frac{1}{6}$ and the system makes a transition to the coplanar-$xy$ state. By looking more precisely to this region we found that indeed this region is composed from two sub-regions of nearly 
coplanar-$xy$ and nearly coplanar-$xz$ states. The nearly coplanar-$xz$ state is situated above the nearly coplanar-$xy$ phase for $J_2$ greater than around $0.30$. For $J_2/J_1 > \frac{1}{2}$ we found a degenerate spiral state with a closed manifolds of stable wave vectors centered at the corners of the Brillouin zone. Again this state remains stable by turning on $g_2$ but it becomes non-planar until the system makes a transition to the collinear-$z$ state. We also found that the manifold of degenerate energy region in the Brillouin zone only depends on $J_2$ and is independent of $g_2$.   

%The spin states in this region resembles similar properties of the known Kawamura states which are some kind of spin configurations combined with several spin wave vectors %\cite{Kawamura}.

\section{Linear Spin Wave Analysis}
In this section we use the linear spin-wave (LSW) theory to 
calculate the zero point quantum corrections to the classical ground state energy.
Moreover,  This method  enables us to investigate
 the stability of a classical state against the quantum fluctuations,
by calculating the  excitation  spectrum.  
For a given classical spin configuration,  first it is convenient  to apply an appropriate  local rotation on each site,  
in order  to turn the local $ z $-axis along the spin direction on that site\cite{Rastelli,Miyake,Singh}. 
 This can be done by the  following transformation
\begin{equation}
\begin{pmatrix} {S}^{\alpha}_{i,x}  \\ {S}^{\alpha}_{i,y} \\ {S}^{\alpha}_{i,z} \end{pmatrix}  =
\begin{pmatrix} \cos \theta^{\alpha}_{i}\cos \phi^{\alpha}_{i} & -
\sin \phi^{\alpha}_{i} & \sin \theta^{\alpha}_{i} \cos \phi^{\alpha}_{i} \\
 \cos \theta^{\alpha}_{i} \sin \phi^{\alpha}_{i} &
\cos \phi^{\alpha}_{i} & \sin \theta^{\alpha}_{i} \sin \phi^{\alpha}_{i} \\
 -\sin \theta^{\alpha}_{i} & 0 & \cos \theta^{\alpha}_{i} \end{pmatrix}
\begin{pmatrix} \tilde{S}^{\alpha}_{i,x} \\ \tilde{S}^{\alpha}_{i,y} \\ \tilde{S}^{\alpha}_{i,z} \end{pmatrix},
\label{rotation_matrix} 
\end{equation}
in which $ \theta^{\alpha}_{i} $ and $\phi^{\alpha}_{i}$ are the polar and azimuthal  angles determining the direction of the spin at site
$ i $ and,  $ \alpha=1,2 $ denotes the sub-lattice index  and 
$ {\tilde S}^{\alpha}_{x,y,z} $ are the  components of the spins in the locally rotated coordinates.
  Then using the linearized Holstein-Primakoff\cite{HP,Anderson,Auerbach} transformations, one can
derive a quadratic bosonic Hamiltonian, whose spectrum gives the magnon energy dispersion.
The linerized Holstein-Primakoff transformations are as:
\begin{equation}
%\nonumber
\begin{array}{cc}
\left\{
\begin{array}{l}
{\tilde S}_{+,i}^{1}\approx {a}_{i}\sqrt{2S} \\
{\tilde S}_{-,i}^{1}\approx {a}_{i}^{\dagger}\sqrt{2S} \\
{\tilde S}_{z,i}^{1}=S-{a}_{i}^{\dagger}{a}_{ i} \\
\end{array}
\right.
&
\left\{
\begin{array}{l}
{\tilde S}_{+,i}^{2}\approx {b}_{ i}^{\dagger}\sqrt{2S} \\
{\tilde S}_{-,i}^{2}\approx {b}_{ i}\sqrt{2S} \\
{\tilde S}_{z,i}^{2}= -S+{b}_{ i}^{\dagger}{b}_{ i}. \\
\end{array}
\right.
\end{array}
\end{equation}
in which $ S $ is the size of the spin, and in writing the above transformation 
we applied the $ \pi $ phase 
difference between the two sub-lattices.  
Substituting the above relations  
 in the $ {\bf KMH} $ Hamiltonian and retaining only the quadratic terms,    
one  obtains   a Hamiltonian of the following form  in LSW approximation
\begin{equation}
\begin{split}
{\cal H}_{\bf {LSW}}& =N\epsilon_{\rm cl}+\frac{S}{2}\sum_{i} { \psi}^{\dag}_{i}{\bf M}_{i}{ \psi}_{i}\\
&+\frac{S}{2}\sum_{i,j} { \psi}^{\dag}_{i}{\bf T}_{ij}{\psi}_{j}-\frac{S}{4}{\bf {Tr}}(M),
\label{lsw}
\end{split}
\end{equation}
where $ \epsilon_{\rm cl} $ is the classical energy per spin and 
  $ {\psi}^{\dag}_i=(a_i, b_i, a^{\dag}_i, b^{\dag}_i) $.
The matrices $ {\bf M}  $ and ${\bf T} $ contain on-site and  the 
 hopping  terms between the interacting sites, respectively.  
 The explicit expression of the elements of these two matrices are given
 in Appendix.A .  Eqs.(\ref{M1}-\ref{parameters}), show  that 
 these elements  are, in general, functions
 of  in-plane ( ${\bf q} $) and 
 normal-to-plane ( ${\bf q}' $)  wave-vectors.
 Calculation of the matrix elements at the  wave-vectors minimizing  
the classical energy,  already found in Sec.III,  and then the  
diagonalization of the $ {\bf {LSW}} $ Hamiltonian, Eq.(\ref{lsw}),
gives us  the excitation spectrum.  
   
For the phases I (N{\'e}el-$ xy $), II (collinear-$ z $) and III (helical-$ xy $), it can be shown 
that the $ {\bf {LSW}} $ Hamiltonian, is translational invariant and so  can be 
easily diagonalized, using  Fourier transformation of the bosonic operators.

\subsection{In-plane Phases}
First, we start with the in-plane phases, N{\'e}el-$ xy $ and helical-$ xy $. 
In this case $ \theta^{\alpha}_{i}=\pi/2 $, 
hence the spin rotation transformation, Eq.(\ref{rotation_matrix}), takes the form
\bea
 {S^{\alpha}_{i,x}} &=& -\sin (\phi^{\alpha}_{i}) \tilde{S}^{\alpha}_{i,y}+
 \cos(\phi^{\alpha}_{i})\tilde{S}^{\alpha}_{i,z}, \nn \\
 {S^{\alpha}_{i,y}} &=& \cos(\phi^{\alpha}_{i})\tilde{S}^{\alpha}_{i,y}+
 \sin(\phi^{\alpha}_{i})\tilde{S}^{\alpha}_{i,z},  \nn \\
 {S^{\alpha}_{i,z}} &=& -\tilde{S}^{\alpha}_{i,x}, \nn \\
\label{planar-rotation}
\eea
where 
$\phi^{1}_{i}=\bq \cdot \br_i$ and $\phi^{2}_{i}=\bq \cdot \br_i+\varphi$. 
For these states the minimum energy is achieved at $ {\bf q}'=0 $,  then 
one can easily see that all the elements of the 
matrices $ M $ and $ T $ in  Eq.(\ref{lsw}) 
are independent of  position $ {\bf r}_{i} $.  This  translational 
symmetry enables us that  by Fourier transformation  of the bosonic  operators  
\bea
c_{\bk}&=&\frac{1}{\sqrt{N/2}} \sum_{i} a_{i} \exp( {-i{\bf k}\cdot{\bf r}_{i}})\nn\\ 
d_{\bk}&=&\frac{1}{\sqrt{N/2}} \sum_{i} b_{i} \exp ({-i{\bf k}\cdot{\bf r}_{i}}),
\eea
and using the rotation Eq.(\ref{planar-rotation}),  to convert the Hamiltonian Eq.(\ref{lsw})  
 to a quadratic form in $ {\bf K} $-space.
Defining 
$\psi_\bk^{\dagger} = \begin{pmatrix} c^\dagger_{\bk} \ d^\dagger_{\bk} \ c_{-\bk} \ d_{-\bk} \end{pmatrix}$,  
we get
\begin{equation}
{\cal H}_{\bf {LSW}} = N\epsilon_{\rm cl}-NSF+
2S\sum_{\bk>0} \psi_\bk^{\dg} H^\pdg_\bk \psi^\pdg_\bk .
\label{Hqu:eqn}
\end{equation}
where 
$ \epsilon_{\rm cl} $ is given by Eq.(\ref{eq:EclLT}),
for  $ F $ we have 
\be
F= J_1 \sum_{\delta_1}\cos(\bq\cdot\delta_1-\phi)-
2(J_2-g_2)\sum_{\delta_2}\cos( \bq\cdot\delta_2),\nn \\
\ee
and   $H_{\bk}$ is a $ 4\times 4 $ matrix
\be
H_\bk = \begin{pmatrix} A_\bk & B_\bk & C_\bk & D_\bk  \\
B^*_\bk & A_\bk & D^*_\bk & C_\bk  \\ 
C_\bk & D_\bk & A_\bk & B_\bk  \\
D^*_\bk & C_\bk & B^*_\bk & A_\bk.
\label{Mk2}
\end{pmatrix}
\ee
whose elements   are given by
\bea
A_\bk &=& \frac{J_1}{2} \sum_{\delta_1}\cos(\bq\cdot\delta_1-\phi)-
(J_2-g_2)\sum_{\delta_2}\cos( \bq\cdot\delta_2) \nn \\ 
&+& \frac{1}{2}\sum_{\delta_2}[J_2+g_2+(J_2-g_2)\cos(\bq\cdot\delta_2)]\cos (\bk\cdot\delta_2)  \nn\\
B_\bk&=& \frac{J_1}{4}\sum_{\delta_1}[1-\cos(\bq\cdot\delta_1-\phi)]e^{-i\bk\cdot\delta_1} \nn \\
C_\bk &=& \frac{1}{2} \sum_{\delta_2}[J_2+g_2-(J_2-g_2)\cos(\bq\cdot\delta_2)]\cos(\bk\cdot\delta_2) \nn\\
D_\bk &=& \frac{J_1}{4}\sum_{\delta_1}[1+\cos(\bq\cdot\delta_1-\phi)]e^{-i\bk\cdot\delta_1}. \nn \\
%F_\bk &=& J_1 \sum_{\delta_1}\cos(\bq\cdot\delta_1-\phi)-
%2(J_2-g_2)\sum_{\delta_2}\cos( \bq\cdot\delta_2).\nn \\
\label{fk-mk}
\eea
in which  the vectors $ \delta_1 $ and $ \delta_2 $ are defined 
by Eqs.(\ref{delta1}) and (\ref{delta2}).

Diagonalizating  $ H_{\bk} $, we obtained the ground state energy as
\begin{equation}
E_{\rm gs} = N\epsilon_{\rm cl} + E_{Q},
\label{gs-energy}
\end{equation}
in which 
\be
E_{Q}=-NSF-2 S \sum_{\bk>0} {\left( \omega^{+}_{\bk} + \omega^{-}_{\bk} \right) },
\label{q}
\ee
is the quantum correction to the classical energy,  and  
\be
\omega^{\pm}_\bk = \sqrt{\alpha_\bk \pm \beta_\bk},
\label{spectrum}
\ee
are the  eigenvalues of  $ {\bf  M}_{\bk} $, where
\bea
\alpha_\bk &=& A^2_\bk - C^2_\bk + |B_\bk|^2 - |D_\bk|^2, \nn \\
\beta_\bk &=& \sqrt{4 |A_\bk B_\bk -C_\bk D_\bk|^2 + (D_\bk B^*_\bk - B_\bk D^*_\bk)^2}. \nn \\  
\label{coeffs}
\eea

\begin{figure}[t]
\includegraphics[angle=-90, scale=0.6]{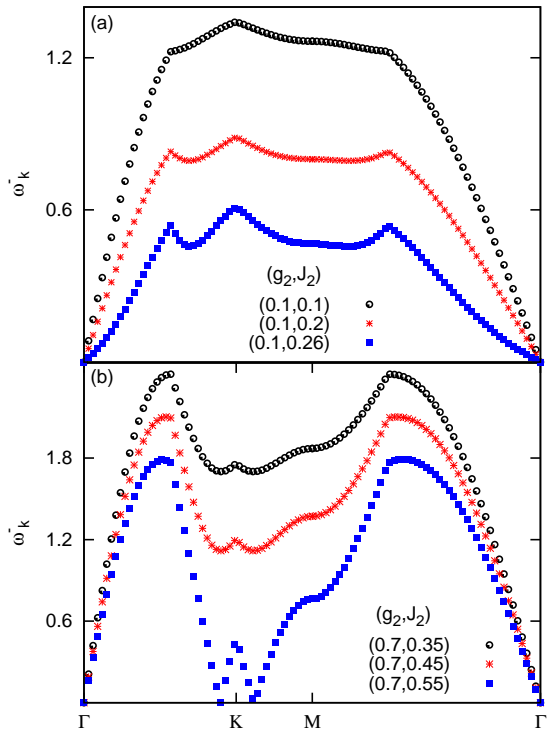}
\caption{The magnon spectrum along symmetry directions in $1BZ$ for phase.I (N{\'e}el-$ xy $). 
a) in going from phase.I to phase.III, b) in going from phase.I to phase.VI. Square symbols show the 
dispersion at the boundaries of this phase.  }
\label{fig:EAF}
\end{figure}

For the phase.I (N{\'e}el-$ xy $), $ {\bq=0} $ and $ \varphi=0 $, 
then from  Eqs.(\ref{fk-mk}), (\ref{spectrum}) and Eq.(\ref{coeffs}),    
one can calculate  the magnon spectrum. 
For some values of the couplings $ J_2 $ and $ g_2 $ within the 
phase.I, the magnon  dispersions are  plotted  along the symmetry directions in $ 1BZ $,
shown  in Fig.~{\ref{fig:EAF}}.  The linear dispersion relation close to the $ \Gamma $-point, in the bulk of this phase
is the result of spontaneous breaking of in-plane $ O_2 $ symmetry leading 
to the formation of the Goldstone modes. The top panel of this figure 
shows that for small value of $ g_2 $, near the phase boundary with helical-$ xy $ (phase.III), the linear
dispersion tends to become quadratic, indicating the appearance  of soft modes which tend to destabilize
the N{\'e}el ordering. On the other hand, the bottom panel,  represents the
emergence of some zero modes at finite wave-vectors around the $ K $-point, at the border of phase.I
and phase.VI. The spectrum becomes non-real in passing from the phase.I to Phase.III
at the $ \Gamma $-point, and at some non-zero wave-vectors  at 
the boundary of phase.I and phase.VI. Therefore, LSW method confirms the classical phase boundaries 
found in the previous section. 
       
\begin{figure}[t]
\includegraphics[angle=-90, scale=0.6 ]{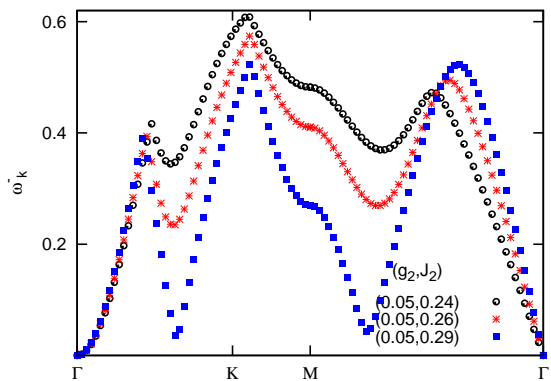}
\caption{The magnon spectrum along symmetry directions in $1BZ$ for phase.III (helical-$ xy $),
in going from phase.III to phase.IV (helical-$ xz $)}
\label{fig:EXS}
\end{figure}

\begin{figure}[t]
\includegraphics[trim=2.5cm 0.5cm 0cm 0cm, clip, scale=0.5]{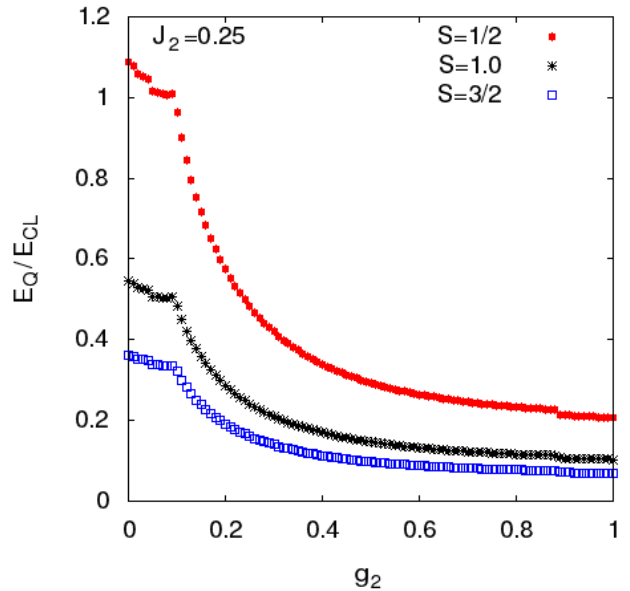}
\caption{The ratio of  quantum correction to 
classical energy ($ E_Q/E_{cl} $) vs $ g_2 $,  for $J_2=0.25 $.
The plots corresponds to  $ S=1/2,1,3/2 $ from top to bottom, respectively. }
\label{quantum-correction}
\end{figure}

In phase.III, we already discussed that the  classical ground state 
 is degenerate  on a  
manifold determined by  Eq.(\ref{eq:spiralJG}). 
However, the LSW analysis shows that only 
the spin state with the wave-vector
\bea
&& {\bq^{*}_{x}}=0\nn\\
&& {\bq^{*}_{y}}=\frac{2}{\sqrt{3}a}\cos^{-1}\left({\frac{{J_1}^2}{16(J_2-g_2)^2}}-\frac{5}{4}\right)
\eea
and the other  five symmetrically equivalent  by $ \pi/6 $ rotations 
(displayed by filled circles in Fig.~{\ref{fig:frustrate}}),
 are stable against the quantum fluctuations, in a sense that their magnon spectrum are real over all $ 1BZ $. 
This is the manifestation of the quantum order-by-disorder, 
already been obtained for $ J_1-J_2 $ model~\cite{aron2010}.   
Indeed, the quantum fluctuations  select only the  
 helical  states along  the nearest-neighbor directions.
   Fig.~\ref{fig:EXS}, exhibits the excitation dispersion
 in the bulk of phase.III for three sets of couplings $ (g_2,J_2) $. 
 The nonlinear dispersion near the $ \Gamma $-points 
 indicates the presence of the soft Goldstone modes in this phase. Close to the boundary with phase.IV, 
 some zero modes tend to emerge which eventually  destabilize the in-plane states.      

In Fig.~{\ref{quantum-correction}}, the ratio of the 
zero point energy ($ E_Q $) to the classical energy ($ E_{cl} $), 
at fixed $ J_{2}=0.25 $, is depicted vs $ g_2 $
 for the  spin sizes $ S=1/2,1,3/2 $. 
This figure shows the increasing effect of quantum fluctuations
when the spin size decreases. For $ S=1/2 $, in the helical-$ xy $ phase, 
the quantum correction is in the order of the classical energy, 
hence it is possible that the helical states melt into some
purely quantum states such as nematic staggered dimerized or plaquette valence bond states.
Such a scenario has  been proposed for $ J_1-J_2 $ model and a plaquette ordering was found 
for $ 1/6<J_2<0.3 $ which transforms to a staggered dimerized state  for $ J_2>0.3 $~\cite{mosadeq,pvb2}.

\subsection{Phase.II }

\begin{figure}[t]
\includegraphics[angle=0, scale=0.5]{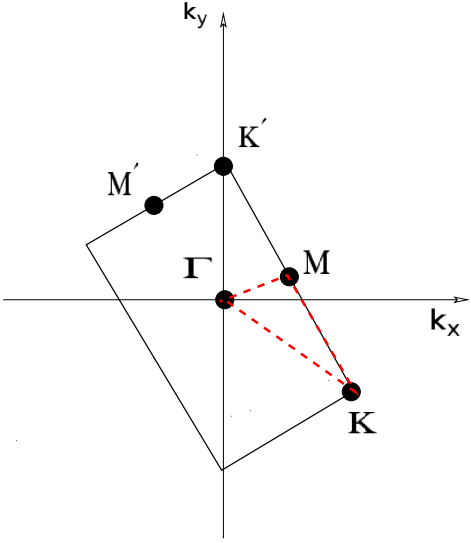}
\caption{The first Brillouin of the magnetic unit cell corresponding to 
 a spin configuration in  phase.II, shown in Fig.~\ref{LT-phase}  }
\label{bz2}
\end{figure}

The magnetic unit cell of a state in phase.II (collinear-$ z $) consists of 
four spins, two  in up  and two in down direction(Fig.~\ref{LT-phase}).
Thus, the  primitive translational vectors for such a unit cell are $ 2{\bf a} $ and $ {\bf b} $, 
and the corresponding first Brillouin zone is a rectangle rotated by $ \pi/4 $ ( Fig.~\ref{bz2}).    
Here, we need four sets of bosonic operators for LSW analysis, as the following  

\begin{figure}[t]
\includegraphics[angle=-90, scale=0.6]{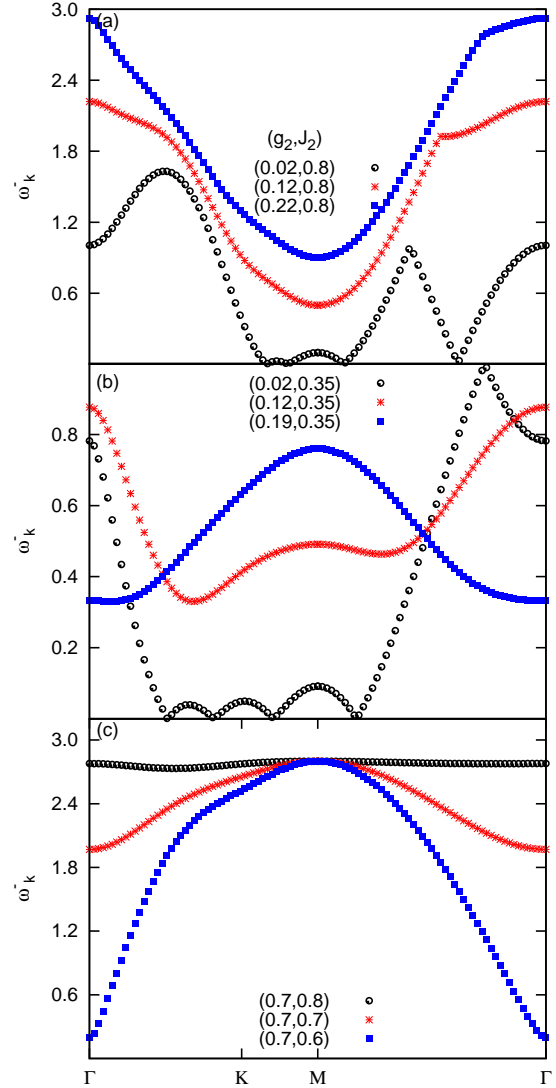}
\caption{The magnon spectrum along symmetry directions in $1BZ$ for phase.II (Collinear-$ z $). 
a) in going from phase.II to phase.V, 
b) in going from phase.II to phase.IV.
c) in going from phase.II to phase.VI}
\label{fig:EXSB1}
\end{figure}
   
\begin{equation}
\nonumber
\begin{array}{cc}
\left\{
\begin{array}{l}
S_{\nu,i}^+\approx\hat{a}_{\nu, i}\sqrt{2S} \\
S_{\nu,i}^-\approx\hat{a}_{\nu, i}^{\dagger}\sqrt{2S} \\
S_{\nu,i}^z=S-\hat{a}_{\nu, i}^{\dagger}\hat{a}_{\nu, i} \\
\nu=1,3
\end{array}
\right.
&
~~~~
\left\{
\begin{array}{l}
S_{\nu,i}^+\approx\hat{a}_{\nu, i}^{\dagger}\sqrt{2S} \\
S_{\nu,i}^-\approx\hat{a}_{\nu, i}\sqrt{2S} \\
S_{\nu,i}^z=\hat{a}_{\nu, i}^{\dagger}\hat{a}_{\nu, i}-S \\
\nu=2,4
\end{array}
\right.
\end{array}
\end{equation}

Using the above transformations, in the linear
approximation we obtain the    following 
quadratic  Hamiltonian,
\begin{equation}
H_{\bf LSW} = E_{\rm cl}+S\sum_{\bk>0} [ \hat{\psi}_\bk^\dag H^\pdg_\bk \hat{\psi}_\bk-G].
\label{HquCZ:eqn}
\end{equation}
in which 
\bea
\hat{\psi}_\textbf{k}^\dag = \begin{pmatrix}
\hat{a}_{1\textbf{k}}^{\dag}~\hat{a}_{2\textbf{k}}^{\dag}~ 
\hat{a}_{3\textbf{k}}^{\dag}~
\hat{a}_{4\textbf{k}}^{\dag}~

\hat{a}_{1-\textbf{k}}~
\hat{a}_{2-\textbf{k}}~
\hat{a}_{3-\textbf{k}}~
\hat{a}_{4-\textbf{k}} \\
\end{pmatrix}, 
%\end{array}
\eea
for the   classical energy  we have  
\bea
E_{\rm cl} &=& -NS^2 (\frac{J_1}{2}+J_2+g_2),
\label{eq:Ecl-spin-wave=CZ}
\eea
and  $ G $ is given by 
\be
G =+4J_1+8(J_2+g_2).
\ee
 $ H_{\bf k} $ is a $ 8\times 8 $ matrix 
\be
H_\textbf{k}=
\left(
\begin{array}{cccccccc}
D_\bk & 0 & 0 & J_1 & 0 & C_\textbf{k}^* & A_\textbf{k}^* & 0 \\
0 &  D_\bk & J_1 & 0 & C_\bk & 0 & 0 & B_\textbf{k}^*  \\
0 & J_1 & D_\textbf{k} & 0 & A_\textbf{k} & 0 & 0 & C_\textbf{k} \\
J_1 & 0 & 0 & D_\textbf{k} & 0 & B_\textbf{k} & C_\textbf{k}^* & 0 \\
0 & C_\textbf{k}^* & A_\textbf{k}^* & 0 & D_\bk & 0 & 0 & J_1 \\
C_\textbf{k} & 0 & 0 & B_\textbf{k}^* & 0 & D_\bk & J_1 & 0 \\
A_\textbf{k} & 0 & 0 & C_\textbf{k} & 0 & J_1 &  D_\bk & 0 \\
0 & B_\textbf{k} & C_\bk^* & 0 & J_1 & 0 & 0 & D_\bk, \\
\end{array}
\right),
\label{Mk3}
\ee
whose elements are  as the following 
\bea
A_\bk &=& J_1(e^{ik_a}+e^{i(k_a+k_b)}) \nn \\
B_\bk &=& J_1(1+e^{ik_b})  \nn \\
C_\bk &=& (J_2-g_2)(1+e^{ik_a}+e^{-ik_b}+e^{i(k_a+k_b)}) \nn \\
D_\bk &=& J_1+2(J_2+g_2)+2(J_2-g_2)\cos(k_b)  \nn \\
\eea
in which $ k_a={2\bf k\cdot \bf a} $ and $ k_b={\bf k\cdot \bf b} $.

Diagonalizing  the matrix $H_\bk $, 
for the lowest    magnon dispersion, we get  

\be
\omega^{-}_\bk = \frac{1}{2}\sqrt{\alpha_\bk - 2\beta_\bk}
\ee
in which 
\bea
\alpha_\bk &=& 2(2 - |B_\bk|^2 -  |A_\bk|^2 - 2 |C_\bk|^2 + 2 {D_\bk}^2) \nn  \\
\beta_\bk &=& [(A_\bk A^*_\bk - B_\bk B^*_\bk)^2 + 
4 |A_\bk C_\bk^* + B_\bk C_\bk|^2  \nn\\
&-& 4 |A_\bk- B_\bk^*|^2 -8 D_\bk(C_\bk(A_\bk^*+B_\bk)\nn\\
 &+&C_\bk^*(A_\bk+B_\bk^*))+16D_\bk^2]^{1/2}\nn \\
\eea

The magnon dispersion for three sets of the couplings 
$ (g_2, J_2) $ is plotted  in Fig.~\ref{fig:EXSB1}. 
As can be seen from the top and middle panels of 
this figure, the appearance of zero modes at the borders 
 of phase.V  and phase.IV, indicates the instability
of collinear-$ z $ states at those  boundaries.  On the other hand
the bottom panel, shows  the appearance of  Goldstone modes
at the border of phase.II and phase.VI, indicating restoration  of
 in-plane $ O_2 $
symmetry. The absence of  the Goldstone modes at the boundaries with phases 
VI and V would be the   sign  of  the  breaking of all rotational symmetries at these two regions. 

\subsection{Phase.IV and V}

In the phase.IV (helical-$ xz $) and phase.V (Non-coplanar), the $ z $-component
of the spins are incommensurate, i.e $ {\acute{\bf{q}}} $ is not equal to any $ M $-point, hence 
the coefficients of the Hamiltonian Eq.(\ref{lsw}) are site dependent and Fourier transformation
would not be useful to find its eigenvalues.
 For $ g_2=0 $ and $ J_2>0.3 $, it has been shown that 
an incommensurate planar  state would be selected by   quantum order 
by   disorder mechanism~\cite{aron2010}.
 Turning on the $ g_2 $ term, the planar states are confined to be 
on plane normal to the horizontal plane and as explained in Sec.II-B, 
these classical states possess $ O_2 $ rotational symmetry in the vertical  plane. 
However   the quantum fluctuations 
 breaks the $ O_2 $ symmetry, and as a result the translational symmetry 
 of the $ {\bf {LSW}} $ Hamiltonian will be lost. 
The coefficients of the Hamiltonian, Eq.(\ref{lsw}) are quasi-periodic,  
which can be considered as  a  random Hamiltonian 
with a long-range correlated disorder.
Therefore, the localization of the magnons  
would be a possible scenario  provided   
$ g_2 $ term be strong  enough. In this case one expects the freezing of the spins in random directions
and emergence of a spin-glass ground state. 
To investigate the localization of magnon, the exact diagonalization of the 
$ {\bf {LSW} } $ Hamiltonian and also the methods such as level statistics and 
inverse participation ratio are required, and  we leave this study for the future works.
However,  since  $ g_2 $ is small for the range of couplings 
in the phase diagram ($ g_2 \lesssim 0.02  $), it can be speculated that the 
magnon localization is not probable in these two phases and the 
helical states selected due to  quantum order by disorder mechanism are still 
stable. Nevertheless, for $ S=1/2 $, the quantum fluctuations may 
destabilize such a helical state in favor of some valence bond configurations.       
              
\section{Conclusion}
In summary, we obtained the zero temperature phase diagram of the 
classical Kane-Mele-Heisenberg model. The competition among   
the isotropic exchange interactions between first $ J_1 $ and second neighbors $ J_2 $,
as well as the second neighbor anisotropic  exchange $ g_2 $,  
gives rise to a rich phase diagram. In the region of the couplings space where all the coupling are positive,
we found six distinct phases. Three phases are long-range ordered, 
namely,  in-plane N{\'e}el (phase.I), commensurate vertical planar (phase.VI) and 
vertical collinear (phase.II).  The other three, being infinitely degenerate
due to the frustrating competition between the couplings, are 
incommensurate  in-plane helical (phase.III), incommensurate vertical planar (phase.VI) 
and incommensurate non-coplanar (phase.V). 
The linear spin wave analysis,  done 
analytically in  phase.III, shows that the quantum order-by-disorder
is at work  in this phase, and a set of rotationally equivalent helical state    
are selected from the ground state manifold, by the quantum fluctuations.
 In the phases IV and V, 
due the lack of translational symmetry in the LSW Hamiltonian, 
the analytic calculations is not possible and numeric exact diagonalization    
of the Hamiltonian is  required to figure out how the quantum fluctuations 
affect the classical ground states in these regions.  This is the subject
of our current research. 

It is also found that for $ S=1/2 $, the quantum correction to the ground state energy is 
at the same order as the  classical energy, hence the selected helical states are likely 
to  melt  down into some purely quantum ground state.    
Whether the emergent phase is a QSL or a kind of plaquette valence bond or else,
one needs to employ the methods such as  exact diagonalization, variational Monte Calro, 
bond operator formalism which may shed more light on the nature of  the quantum ground state
in the phase as well as the phases IV and V.  

 As the final remark, we  add that the methods employed in this paper can be used for obtaining the phase diagram of the spin systems in which 
the axial spin symmetry is also broken. Such models are realized in monolayers of Sodium Iridate  ($Na_2IrO_3$ ), where the next nearest hopping term 
in its tight-binding model is link dependent due to dependence of intrinsic spin-orbit coupling to the hopping direction~\cite{shitade}.
The spin Hamiltonian arising in the strong coupling limit of this model is recently investigated and shown to have a rich phase diagram~\cite{Rachel,Kargarian}.     
The complete breaking of the spin $SU(2)$ symmetry  also occurs when  a Rashba spin-orbit coupling is added to the KM model,  leading  to 
a link dependent exchange interaction in the large coupling limit~\cite{Rachel}. The classical phase diagrams of aforementioned  models are 
currently under investigation in our group.

%\acknowledgments

%\bigskip

%\newpage
\appendix

\section{Matrix elements of LSW Hamiltonian in real space }

The matrices introduced  the spin-wave Hamiltonian Eq.(\ref{lsw}) are defined as

\bea
M_i = \begin{pmatrix} \epsilon^{i}_1 & 0 & 0 & 0\\
0 & \epsilon^{i}_2 & 0 & 0 \\ 
0 & 0 & \epsilon^{i}_1 & 0\\
0 & 0 & 0 & \epsilon^{i}_2
\label{M1}
\end{pmatrix}
\eea

and

\bea
T_{ij} = \begin{pmatrix} {t^{ij}_1} & {t^{ij}_5} & {t^{ij}_3} & {t^{ij}_6}  \\
{t^{ij}_5}^{*} & {{t^{ij}_2}} & {{t^{ij}_6}}^{*} & {{t^{ij}_4}}   \\ 
{{t^{ij}_3}}^{*} & {{t^{ij}_6}}^{*} & {{t^{ij}_1}}^{*} & {{t^{ij}_5}}^{*}  \\
{{t^{ij}_6}} & {{t^{ij}_4}}^{*} & {{t^{ij}_5}} & {{{t^{ij}_2}}^{*}},
\label{T1}
\end{pmatrix}
\eea
whose elements are given by 
\bea
\epsilon^{i}_1&=&-2\chi_3+{t^{ij}_7}\nn \\
\epsilon^{i}_2&=&-2\chi_4+{t^{ij}_7}\nn \\
{t^{ij}_1}&=&\chi_1+\chi_5+i\chi_6-i\chi_7\nn \\
{t^{ij}_2}&=&\chi_2+\chi_5-i\chi_8+i\chi_9\nn \\
{t^{ij}_3}&=&\chi_1-\chi_5-i\chi_6-i\chi_7\nn \\
{t^{ij}_4}&=&\chi_2-\chi_5+i\chi_8+i\chi_9\nn \\
{t^{ij}_5}&=&\frac{J_1}{2}[(\zeta_1-1)\cos(\bq\cdot\delta_1-\phi)+\zeta_2-i\zeta_3-i\zeta_4]\nn \\
{t^{ij}_6}&=&\frac{J_1}{2}[(\zeta_1+1)\cos(\bq\cdot\delta_1-\phi)+\zeta_2+i\zeta_3-i\zeta_4]\nn \\
{t^{ij}_7}&=&J_1[\zeta_1+\zeta_2\cos(\bq\cdot\delta_1-\phi)],\nn \\
\label{tm-elements}
\eea
where
\bea
\zeta_1&=&\frac{1}{2}[\cos(\acute{\bq}\cdot\delta_1+\gamma-\eta)\nn \\
&+&\cos(2\acute{\bq}\cdot\br_i+\acute{\bq}\cdot\delta_1+\gamma+\eta)]\nn \\
\zeta_2&=&\frac{1}{2}[\cos(\acute{\bq}\cdot\delta_1+\gamma-\eta)\nn \\
&-&\cos(2\acute{\bq}\cdot\br_i+\acute{\bq}\cdot\delta_1+\gamma+\eta)]\nn \\
\zeta_3&=&\sin(\bq\cdot\delta_1-\phi)\cos(\acute{\bq}\cdot\br_i+\gamma)\nn \\
\zeta_4&=-&\sin(\bq\cdot\delta_1-\phi)\cos(\acute{\bq}\cdot\br_i+\acute{\bq}\cdot\delta_1+\eta)\nn \\
\eea

\bea
\chi_1&=&\frac{(J_2-g_2)}{2}\cos(\bq\cdot\delta_2)[\cos(\acute{\bq}\cdot\delta_2)\nn \\
&+&\cos(2\acute{\bq}\cdot\br_i+\acute{\bq}\cdot\delta_2+2\gamma)]\nn \\
&+&\frac{(J_2+g_2)}{2}[\cos(\acute{\bq}\cdot\delta_2)-\cos(2\acute{\bq}\cdot\br_i+\acute{\bq}\cdot\delta_2+2\gamma)]\nn \\
\chi_2&=&\frac{(J_2-g_2)}{2}\cos(\bq\cdot\delta_2)[\cos(\acute{\bq}\cdot\delta_2)\nn \\
&+&\cos(2\acute{\bq}\cdot\br_i+\acute{\bq}\cdot\delta_2+2\eta)]\nn \\
&+&\frac{(J_2+g_2)}{2}[\cos(\acute{\bq}\cdot\delta_2)-\cos(2\acute{\bq}\cdot\br_i+\acute{\bq}\cdot\delta_2+2\eta)]\nn \\
\chi_3&=&\frac{(J_2-g_2)}{2}\cos(\bq\cdot\delta_2)[\cos(\acute{\bq}\cdot\delta_2)\nn \\
&-&\cos(2\acute{\bq}\cdot\br_i+\acute{\bq}\cdot\delta_2+2\gamma)]\nn \\
&+&\frac{(J_2+g_2)}{2}[\cos(\acute{\bq}\cdot\delta_2)+\cos(2\acute{\bq}\cdot\br_i+\acute{\bq}\cdot\delta_2+2\gamma)]\nn \\
\chi_4&=&\frac{(J_2-g_2)}{2}\cos(\bq\cdot\delta_2)[\cos(\acute{\bq}\cdot\delta_2)\nn \\
&-&\cos(2\acute{\bq}\cdot\br_i+\acute{\bq}\cdot\delta_2+2\eta)]\nn \\
&+&\frac{(J_2+g_2)}{2}[\cos(\acute{\bq}\cdot\delta_2)+\cos(2\acute{\bq}\cdot\br_i+\acute{\bq}\cdot\delta_2+2\eta)]\nn \\
\chi_5&=&(J_2-g_2)\cos(\bq\cdot\delta_2)\nn \\
\chi_6&=&(J_2-g_2)\sin(\bq\cdot\delta_2)\cos(\acute{\bq}\cdot\br_i+\gamma)\nn \\
\chi_7&=-&(J_2-g_2)\sin(\bq\cdot\delta_2)\cos(\acute{\bq}\cdot\br_i+\acute{\bq}\cdot\delta_2+\gamma)\nn \\
\chi_8&=&(J_2-g_2)\sin(\bq\cdot\delta_2)\cos(\acute{\bq}\cdot\br_i+\eta)\nn \\
\chi_9&=-&(J_2-g_2)\sin(\bq\cdot\delta_2)\cos(\acute{\bq}\cdot\br_i+\acute{\bq}\cdot\delta_2+\eta)\nn \\
\label{parameters}
\eea

%In the above expressions  $\delta_1={\bf 0}, \pm\textbf{b}, \pm(\textbf{a+b})$
% with $ - $ for the first sub-lattice  and $ + $  for the second one,  and 
%$\delta_2=\pm\textbf{a}, \pm\textbf{b}, \pm(\textbf{a+b})$ are the position vectors  of the 
% second neighbor primitive cells.

\end{document}